\documentclass{pasa}%

\usepackage{graphicx}
\usepackage{xcolor}

\usepackage[]{hyperref}
\hypersetup{
    colorlinks,
    linkcolor={red!50!black},
    citecolor={blue!50!black},
    urlcolor={blue!80!black}
}

\newcommand{\gw}{{\rm gw}}

\def\xlinkspace#1 #2{%
 \ifx\relax#2%
 \xlinkdash#1-\relax
 \else
 \xlinkdash#1 -\relax
 \expandafter\xlinkspace\expandafter#2%
 \fi}

\def\xlinkdash#1-#2{%
 \ifx\relax#2%
 \tmp{#1}%
 \else
 \tmp{#1-}%
 \expandafter\xlinkdash\expandafter#2%
 \fi}

\title[Gravitational waves from neutron stars]{Gravitational Waves from Neutron Stars: A Review}
\author[Lasky]{Paul D. Lasky$^{1}$\thanks{email: paul.lasky@monash.edu}
\\
\affil{$^1$Monash Centre for Astrophysics, School of Physics and Astronomy, Monash University, VIC 3800, Australia}
}%
\jid{PASA}
\doi{10.1017/pas.\the\year.xxx}
\jyear{\the\year}

\usepackage[authoryear]{natbib}
\bibpunct{(}{)}{;}{a}{}{,}
\begin{document}%
\begin{abstract}
Neutron stars are excellent emitters of gravitational waves.
Squeezing matter beyond nuclear densities invites exotic physical processes, many of which violently transfer large amounts of mass at relativistic velocities, disrupting spacetime and generating copious quantities of gravitational radiation.  I review mechanisms for generating gravitational waves with neutron stars.  This includes gravitational waves from radio and millisecond pulsars, magnetars, accreting systems and newly born neutron stars, with mechanisms including magnetic and thermoelastic deformations, various stellar oscillation modes and core superfluid turbulence.  I also focus on what physics can be learnt from a gravitational wave detection, and where additional research is required to fully understand the dominant physical processes at play.
\end{abstract}
\maketitle%

\section{Introduction}\label{sec:intro}
The dawn of gravitational wave astronomy is one of the most anticipated scientific advances of the coming decade.  The second generation, ground-based gravitational wave interferometers, Advanced LIGO \citep[aLIGO][]{ligo15} and Virgo \citep{acernese15}, are due to start observing later in 2015 and in 2016, respectively.  The first aLIGO observing run is expected to 
be a few times more sensitive than
initial LIGO, with a full order of magnitude increase in strain sensitivity by $\sim2019$ \citep[for a review, see][]{ligo13}.  

The inspiral and merger of compact binary systems (neutron stars and/or black holes) are commonly expected to be the first detections with aLIGO.  The most robust predictions for event rates come from observations of the binary neutron star population within our galaxy, with an expected binary neutron-star detection rate for the aLIGO/Virgo network at full sensitivity between 0.4 and 400 per year \citep{abadie10}.  But there are many other exciting astrophysical and cosmological sources of gravitational waves \citep[for a brief review, see][]{riles13}.  Loosely, these can be divided into four categories: compact binary coalescences, bursts, stochastic backgrounds and continuous waves.  Burst sources include gravitational waves generated in nearby supernova explosions, magnetar flares, and cosmic string cusps.  Stochastic backgrounds arise from the incoherent sum of sources throughout the Universe, including from compact binary systems, rotating neutron stars, and primordial perturbations during inflation.  Continuous gravitational waves are almost monochromatic signals generated typically by rotating, non-axisymmetric neutron stars.  

The above laundry list of gravitational wave sources prominently features neutron stars in their many guises.  While supranuclear densities, relativistic velocities and enormous magnetic fields are exactly what makes neutron stars amenable to emitting gravitational waves of sufficient amplitude to be detectable on Earth, it is also these qualities that makes it difficult to provide accurate predictions of the gravitational wave amplitudes, and hence detectability, of their signals.  A positive gravitational wave detection from a neutron star would engender great excitement, but it is the potential to understand the interior structure of neutron stars that will make this field truly revolutionary.

In this review, I provide a detailed overview of many proposed gravitational wave generation mechanisms in neutron stars, including state-of-the-art estimates of the gravitational wave detectability.  These include gravitational waves generated from magnetic deformations in newly-born and older isolated radio pulsars (section \ref{mag_def_sec}), accreting systems (section \ref{accrete_sec}), impulsive and continuous-wave emission from pulsar glitches (section \ref{glitch_sec}), magnetar flares (section \ref{flare_sec}), and from superfluid turbulence in the stellar cores (section \ref{turb_sec}).  As well as discussing gravitational wave detectability, I also concentrate on what physics can be learnt from a future, positive gravitational wave detection.

It is worth stressing that this is {\it not} a review of all gravitational wave sources in the audio band, but is instead designed to review the theory of gravitational wave sources from neutron stars.  The article is but one in a series highlighting Australia's contribution to gravitational wave research \citep{howell15,kerr15,slagmolen15}.




\section{Magnetic Deformations}\label{mag_def_sec}
Spinning neutron stars possessing asymmetric deformations emit gravitational waves.  Such deformations can be generated through elastic strains in the crust \citep{bildsten98,ushomirsky00}, or strong magnetic fields in the core \citep{bonazzola96}.  A triaxial body rotating about one of it's principal moments of inertia will emit radiation at frequency $f_\gw=2\nu$, where $\nu$ is the star's spin frequency.  More precisely, a freely precessing, axisymmetric body, with principal moment of inertia, $I_{\rm zz}$, and equatorial ellipticity, $\epsilon$, emits a characteristic gravitational wave strain \citep{zimmerman79}
\begin{align}
	h_0&=\frac{4\pi^2G}{c^4}\frac{I_{\rm zz}f_\gw^2\epsilon}{d}\notag\\
	&=4.2\times10^{-26}\left(\frac{\epsilon}{10^{-6}}\right)\left(\frac{P}{10\,{\rm ms}}\right)^{-2}\left(\frac{d}{1\,{\rm kpc}}\right)^{-1},
	\label{h0}
\end{align}
where $d$ is the distance to the source.  For comparison, the smallest, upper limit on stellar ellipticity for young neutron stars in supernova remnants comes from LIGO observations of Vela Jr. (G266.2-1.2), with $\epsilon\le2.3\times10^{-7}$ \citep{aasi14a}.  The overall ellipticity record-holder is for the millisecond pulsars J2124--3358 and J2129--5721 with $\epsilon\le6.7\times10^{-8}$ and $\epsilon\le6.8\times10^{-8}$, respectively \citep{aasi14b}.  Typical ellipticity constraints for isolated radio pulsars are $\epsilon\lesssim10^4$ -- $10^6$ \citep{aasi14b}.

\subsection{Spin down limit}
\begin{figure*}
\includegraphics[width=1.0\columnwidth]{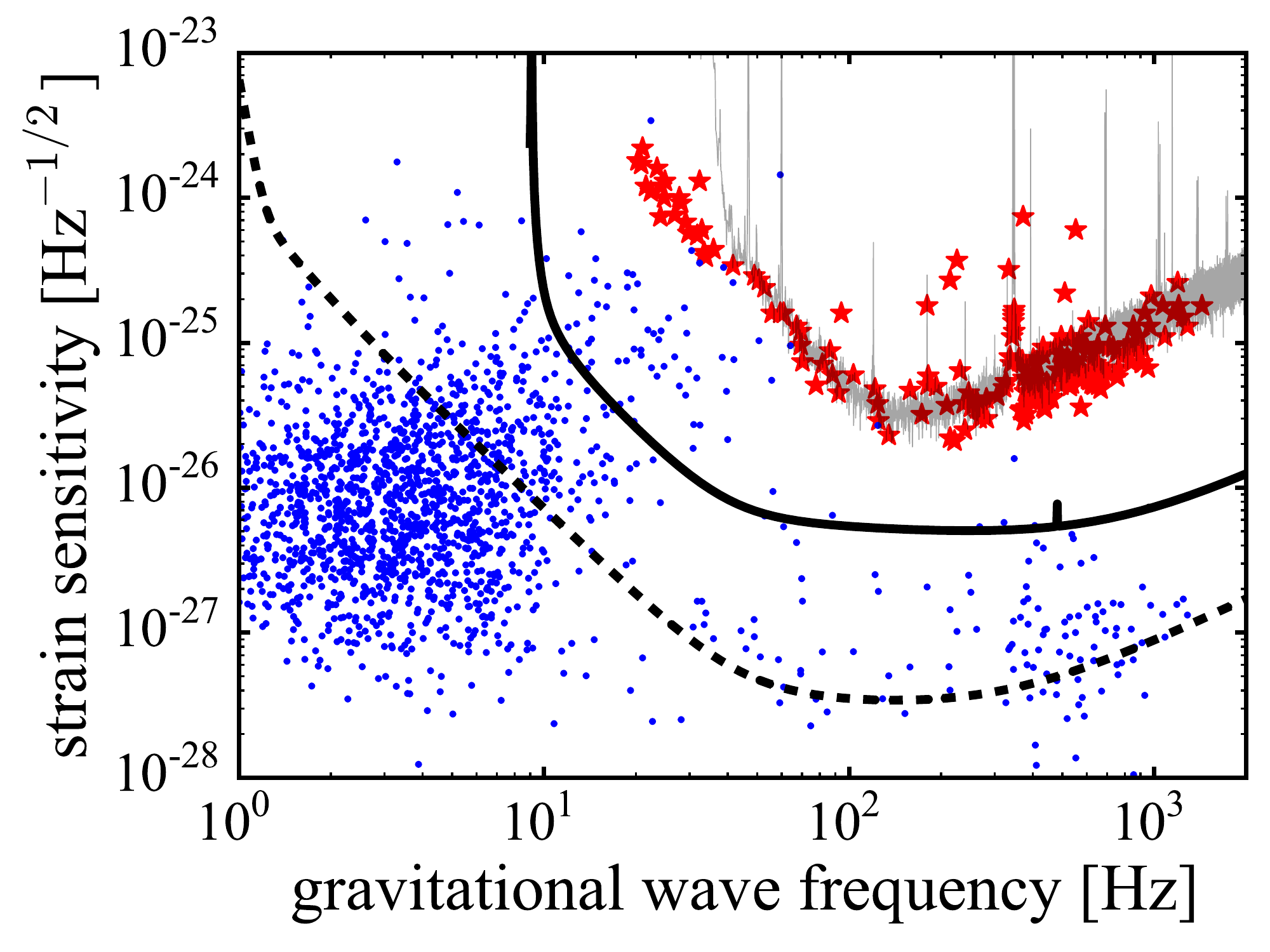}
\includegraphics[width=1.0\columnwidth]{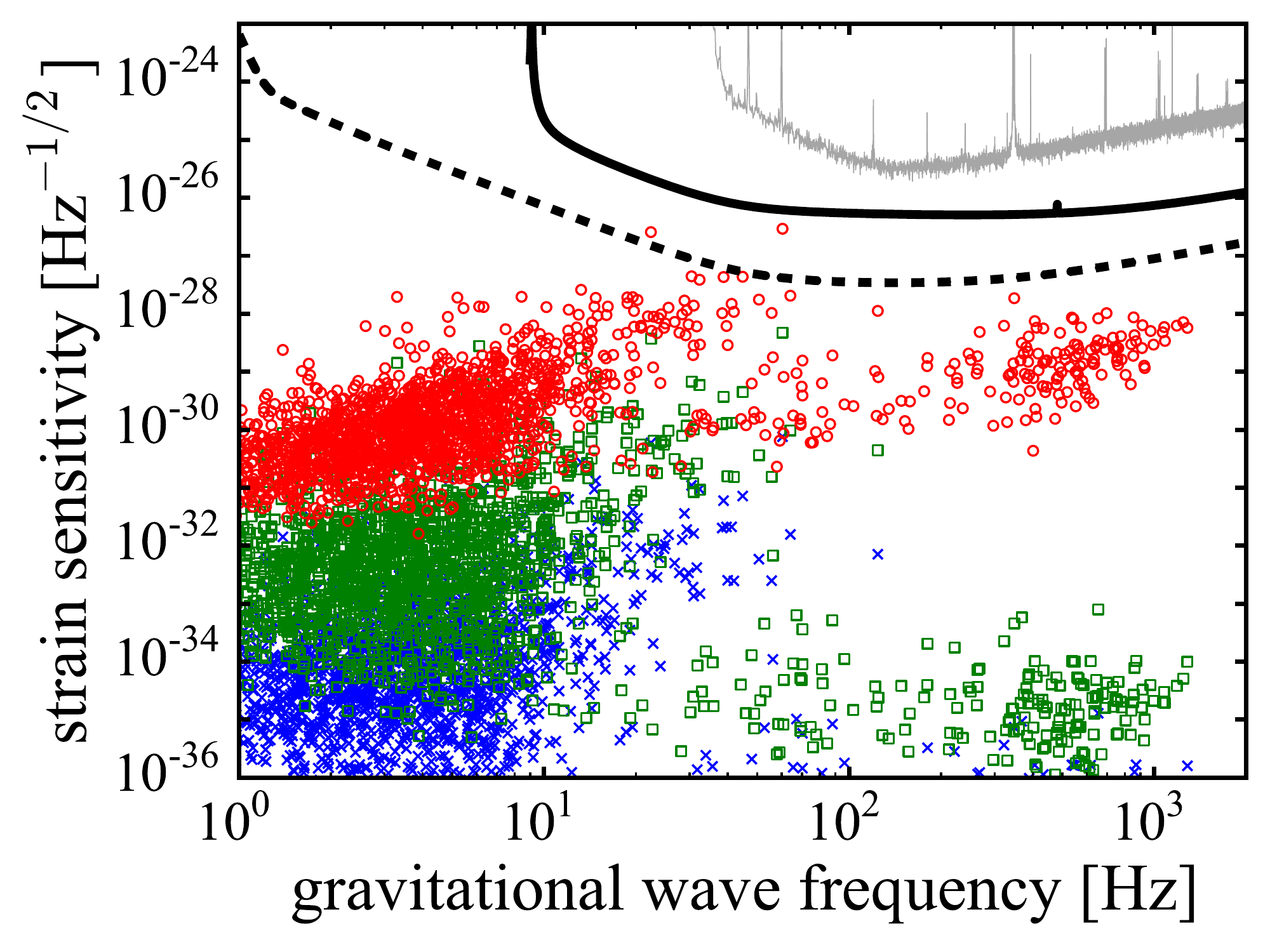}
	\caption{\label{Spindown_ULs} Left panel:  Current upper limits on the gravitational wave strain from known pulsars \citep[red stars; data from][]{aasi14b} and the spin down limits for known pulsars in the ATNF catalogue (blue dots).  Right panel: Gravitational wave strain predictions for known pulsars.  The blue crosses and green squares are for normal neutron star matter with purely poloidal magnetic fields and $\Lambda=0.01$, respectively [see equation (\ref{eps}) and surrounding text].  The red circles assume the neutron stars are colour-flavour-locked phase [CFL; equation (\ref{CFL})] with $\left<B\right>=10B_{\rm p}$.  In both figures, the solid and dashed black curves show the projected strain sensitivity for aLIGO and ET respectively, and the grey curve is the strain sensitivity for the initial S5 run assuming a two-year coherent integration \citep[e.g., see][]{dupuis05}}
\end{figure*}

An absolute upper limit on the gravitational wave strain from individual pulsars, known as the spin down limit, can be calculated assuming the observed loss of rotational energy is all going into gravitational radiation.  The left hand panel of Figure \ref{Spindown_ULs} shows the current upper limits on the gravitational wave strain from a search for known pulsars \citep[red stars;][]{aasi14b}, as well as the spin down limits for the known pulsars in the ATNF catalogue \citep{manchester05}\footnote{\href{http://www.atnf.csiro.au/people/pulsar/psrcat/}{http://www.atnf.csiro.au/people/pulsar/psrcat/}}.  The observed gravitational wave upper limits beat the spin down limit for both the Crab and Vela pulsars, while a further five pulsars are within a factor of five.  Also plotted in Figure \ref{Spindown_ULs} are the projected strain sensitivities for aLIGO and ET (solid and dashed black curves respectively), and the strain sensitivity for  the S5 run of initial LIGO \citep[for details, see][]{aasi14b}.  Age-based upper limits on the gravitational wave strain can also be calculated for neutron stars with unknown spin frequencies \citep{wette08}.

\subsection{Magnetic field-induced ellipticities}
As a general rule of thumb, a neutron star's ellipticity scales with the square of the volume-averaged magnetic field {\it inside} the star, $\left<B\right>$ \citep[e.g.,][]{haskell08}, implying the characteristic gravitational wave strain also scales as $h_0\sim\left<B\right>^2$.  The dipole, poloidal component of the magnetic field at the {\it surface} of the star is inferred from the star's spin period and it's derivative, but very little is known about the interior field strength and/or configuration.  

A considerable body of work has been devoted to understanding possible magnetic field configurations\footnote{In Australia, work on possible magnetic field configurations in the cores of stars dates back to the 1960's, where  \citet{monaghan65,monaghan66b} calculated magnetic field equilibria in polytropes, which even included calculations of magnetic-field induced stellar ellipticities.}, in part to answer the question of gravitational wave detectability, but also to understand the inverse problem regarding what can be learnt from future gravitational wave observations.  Purely poloidal and purely toroidal fields are known to be unstable on dynamical timescales \citep[e.g.,][]{wright73,braithwaite06,braithwaite07}.  Mixed fields, where a toroidal component threads the closed-field-line region of the poloidal field, commonly termed `twisted-torus' fields, are generally believed to be dynamically stable \citep[e.g.,][]{braithwaite06b,ciolfi09,akgun13}, although it has been suggested this is dependent on the equation of state \citep{lander12,mitchell15}.  Moreover, a recent study of the effect twisted-torus fields have on crust-core rotational coupling during neutron star spin down suggest these fields may be unstable on a spin down timescale \citep{glampedakis15}.

While generating solutions to the equations of Newtonian and general relativistic magnetohydrostatics is a noble task, it is no substitute for understanding the initial-value problem that brings one towards realistic magnetic field configurations.  In Newtonian simulations with soft equations of state (typically more applicable to main-sequence stars than to neutron stars), Braithwaite and co. showed magnetic fields evolve either to axisymmetric, twisted-tori \citep{braithwaite06b, braithwaite09}, or non-axisymmetric configurations \citep{braithwaite08}, depending on the initial conditions.  General relativistic magnetohydrodynamic simulations ubiquitously show the development of non-axisymmetries \citep{kiuchi11,lasky11,lasky12,ciolfi11,ciolfi12}, however these simulations begin with a restricted set of initial conditions.  The long-term evolution to stable equilibria in these systems is still an open question.

The aforementioned uncertainty in possible magnetic field configurations translates to uncertainty in gravitational wave predictions.  The relative strengths of poloidal and toroidal components strongly affects the stellar ellipticity, and hence gravitational wave detectability \citep{haskell08,colaiuda08,ciolfi09,mastrano11}.  Relatively standard models suggest \citep[e.g.,][]{mastrano11}
\begin{align}
	\epsilon\approx4.5\times10^{-7}\left(\frac{B_{\rm p}}{10^{14}\,\mbox{G}}\right)^{2}\left(1-\frac{0.389}{\Lambda}\right),
	\label{eps}
\end{align}
where $B_{\rm p}$ is the poloidal component of the surface magnetic field, $\Lambda$ is the ratio of poloidal-to-total magnetic field energy (i.e., $0\le\Lambda\le1$, with $\Lambda=0,\,1$ corresponding to purely toroidal and purely poloidal fields repsectively), and I have normalised to fiducial values of stellar mass and radius.  Unfortunately, equation (\ref{eps}) is pessimistic for gravitational wave detection with aLIGO as the only neutron stars with $B_{\rm p}\sim10^{14}\,\mbox{G}$ are the magnetars, but with spin periods of $P\sim1$ -- $10\,{\rm s}$ they emit gravitational waves at frequencies too low.  The right hand panel of Figure \ref{Spindown_ULs} shows predictions for the gravitational wave strain for known pulsars from the ATNF pulsar catalogue \citep{manchester05}.  The blue crosses assume a purely poloidal field i.e., equation (\ref{eps}) with $\Lambda=1$, while the green squares assume a strong internal toroidal component given by $\Lambda=0.01$.  The red circles are described below, but with normal neutron star matter the prospects for gravitational wave detection from magnetic field-induced non-axisymmetries are grim.  One therefore has to hope that Nature has been kind, and has provided neutron stars with strong internal, toroidal components of the magnetic field.  

\subsection{Generating large ellipticities}
\vspace{0.1cm}
Certainly, one expects {\it newly born} neutron stars to have large internal toroidal fields and correspondingly large ellipticities; strong differential rotation combined with turbulent convection in the nascent neutron star drives an $\alpha$--$\Omega$ dynamo, which winds up a toroidal field as strong as $\sim10^{16}\,\mbox{G}$ \citep[e.g.,][]{duncan92}.  Although the symmetry axis of the wound-up field is aligned to the spin axis of the star, the ellipsoidal star will evolve on a viscous dissipation timescale to become an orthogonal rotator through free-body precession, and hence optimal emitter of gravitational waves \citep{cutler02,stella05,dallosso09}.  How such a strong field evolves over secular timescales is an open question.

A recent spate of papers has shown that strong toroidal fields can also be achieved in systems in dynamical equilibrium by prescribing different forms for the azimuthal currents in the star \citep{ciolfi13}, and also by invoking stratified, two fluid stellar models \citep[i.e., where the neutrons form a superfluid condensate and the protons are either a normal fluid or a superconductor; ][]{glampedakis12a, lander12b}.  Interestingly, exotic states of matter in the core such as crystalline colour-superconductors allow for significantly higher ellipticities \citep{owen05,haskell07}.  Recently, \citet{glampedakis12} showed that, if the ground state of neutron star matter is a colour-superconductor, then the colour-magnetic vortex tension force leads to significantly larger mountains than for normal proton superconductors.  They derived fiducial ellipticities for purely poloidal fields of\footnote{It is worth noting that, in any kind of superconductor, the ellipticity scales linearly with the volume-averaged magnetic field \citep{cutler02}, cf. $\left<B\right>^2$ for normal matter.}
\begin{align}
	\epsilon^{2SC}&\approx4.0\times10^{-6}\frac{\left<B\right>}{10^{14}\,\mbox{G}},\label{CFL}\\
	\epsilon^{CFL}&\approx1.2\times10^{-5}\frac{\left<B\right>}{10^{14}\,\mbox{G}},
\end{align}
where $\epsilon^{2SC}$ and $\epsilon^{CFL}$ respectively denote the ellipticities if matter is in a two-flavour phase (i.e., only the $u$ and $d$ quarks form superconducting pairs) and a colour-flavour-locked phase (where all three quark species are paired), and $\left<B\right>$ is the volume averaged magnetic field in the core of the star.  The red circles in Figure \ref{Spindown_ULs} show gravitational wave predictions for a colour-flavour locked superconductor, equation (\ref{CFL}), with the average internal field 10 times the observationally inferred surface field, i.e., $\left<B\right>=10B_{\rm p}$.  These results are fascinating; \textit{a positive detection of gravitational waves from magnetic deformations in neutron stars is a fundamental probe of the fundamental state of nuclear matter}.

It is worth mentioning that the discussion in this section pertains to the optimal case of a rigidly rotating, axisymmetric body rotating about one of it's principal moments of inertia.  Such a body emits monochromatic gravitational waves at a frequency of $2\nu$.  If the neutron star core contains a pinned superfluid, it will emit also at the spin frequency, $\nu$ \citep{jones10}.  In general,  a {\it non-aligned} rotator will emit at $\nu$, $2\nu$, and a number of frequencies straddling these values \citep{zimmerman80, jones02,vandenbroeck05,lasky13b}.  For radio pulsars, such modulations would also be present in other observables such as pulse time-of-arrivals and radio polarisation \citep{jones01,jones02}.

\subsection{Oscillations in young neutron stars}
The strongest emitters of gravitational waves from magnetic deformations are likely young, rapidly rotating neutron stars.  Such stars may also emit through other channels, the most likely being the unstable $r$-mode, whose restoring force is the Coriolis force; see section \ref{rmode} for a detailed description of the physics of $r$-modes.  Targeted gravitational wave searches of young neutron stars in supernova remnants can be adapted to set limits on the amplitude of such oscillations \citep{owen10}.  This was first done with a 12-day coherent search of S5 LIGO data targeting the neutron star in the supernova remnant Cassiopeia A \citep{wette08,abadie10a}.  The search has since been extended to nine young supernova remnants, with the most sensitive $r$-mode fractional amplitude being less than $4\times10^{-5}$ for Vela Jr. \citep{aasi14a}.  Such a limit is encroaching `interesting values' of the amplitude when compared to simulations that calculate the nonlinear saturation amplitude of various $r$-modes in young neutron stars \citep{bondarescu09,aasi14a}.

\section{Accreting systems}\label{accrete_sec}
\subsection{Torque balance}
An observational conundrum drives research into gravitational wave emission from accreting systems; namely, the absence of accreting pulsars with spin frequencies $\nu\gtrsim700\,{\rm Hz}$ \citep{chakrabarty03,patruno10}.  The argument is simple: measured accretion rates allow one to calculate the angular momentum being transferred to the neutron star, which {\it should} be spun up to frequencies at, or near, the breakup frequency \citep[$\nu\gtrsim1\,{\rm kHz}$ for most equations of state][]{cook94}.  The common explanation is that the neutron stars are losing angular momentum through the emission of gravitational radiation \citep{papaloizou78,wagoner84,bildsten98}, although alternatives exist, most notably invoking interactions between the accretion disk and the companion star's magnetic field \citep[e.g.,][]{white97,patruno12}.  

Balancing the accretion torques with gravitational wave emission allows one to estimate an upper limit on the gravitational wave strain independent of the emission mechanism \citep{wagoner84}:
\begin{align}
	h_0^{\rm EQ}=5.5\times10^{-27}\frac{R_{10}^{3/4}}{M_{1.4}^{1/4}}\left(\frac{F_X}{F_\star}\right)^{1/2}\left(\frac{300\,{\rm Hz}}{\nu}\right)^{1/2},\label{torquebalance}
\end{align}
where $R_{10}=R/10\,{\rm km}$, $M_{1.4}=M/1.4\,M_\odot$, $F_X$ is the X-ray flux and $F_\star=10^{-8}\,{\rm erg\,cm}^{-2}\,{\rm s}^{-1}$.  The $\nu^{-1/2}$ scaling arises only because of the spin-equilibrium assumption; slower rotators need larger ellipticities for torque balance.  If one builds a mountain {\it without} being in spin equilibrium, then the gravitational wave strain is simply given by equation (\ref{h0}).

Equation (\ref{torquebalance}) implies the loudest gravitational wave emitters are the brightest X-ray sources such as the low-mass X-ray binaries, the brightest of which being Scorpius X-1 (Sco X-1).  \citet{watts08} utilised observations of Sco X-1 and other known accreting systems to determine their gravitational wave detectability with current and future interferometers showing that, even at torque balance, most systems will be very difficult to detect with aLIGO.

The LIGO Scientific Collaboration has given periodic gravitational waves from Sco X-1 high priority, with multiple data analysis pipelines already fully developed \citep[see][and references therein]{messenger15}.  The unknown spin period of the neutron star in Sco X-1 and the variable accretion rate that drives spin-wandering of the neutron star complicate these searches, but the best gravitational wave upper limit utilises the Sideband search \citep{sammut14}, which gives a 95\% upper limit of $h_0\lesssim8\times10^{-25}\,{\rm Hz}$ at 150 Hz \citep{aasi15}.  This is still above the torque balance limit for Sco X-1, $h_0^{\rm EQ}\approx3.5\times10^{-26}\left(300\,{\rm Hz}/\nu\right)^{1/2}$, but this is expected to be beaten with aLIGO observations \citep{whelan15}.  

Torque balance is an empirically derived limit that is independent of the mechanism generating the gravitational radiation.  From a theoretical perspective, there are a number of ways in which sufficient energy can be lost to gravitational waves, including unstable oscillation modes \citep{andersson98a}, and non-axisymmetric deformations supported by the magnetic field \citep{cutler02,melatos05} or elastic crust \citep{bildsten98,ushomirsky00}.

\begin{figure*}
\includegraphics[width=1.0\columnwidth]{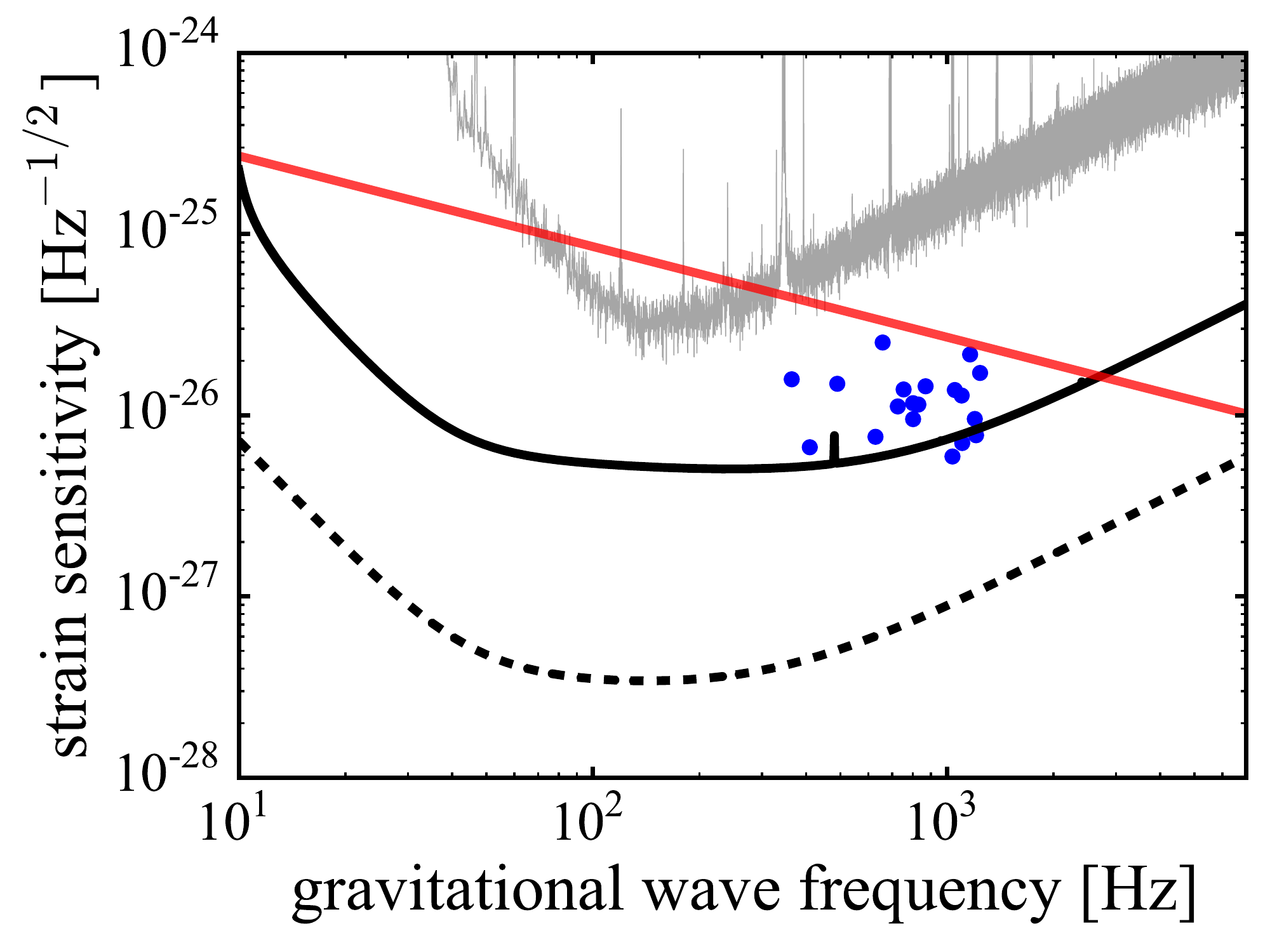}
\includegraphics[width=1.0\columnwidth]{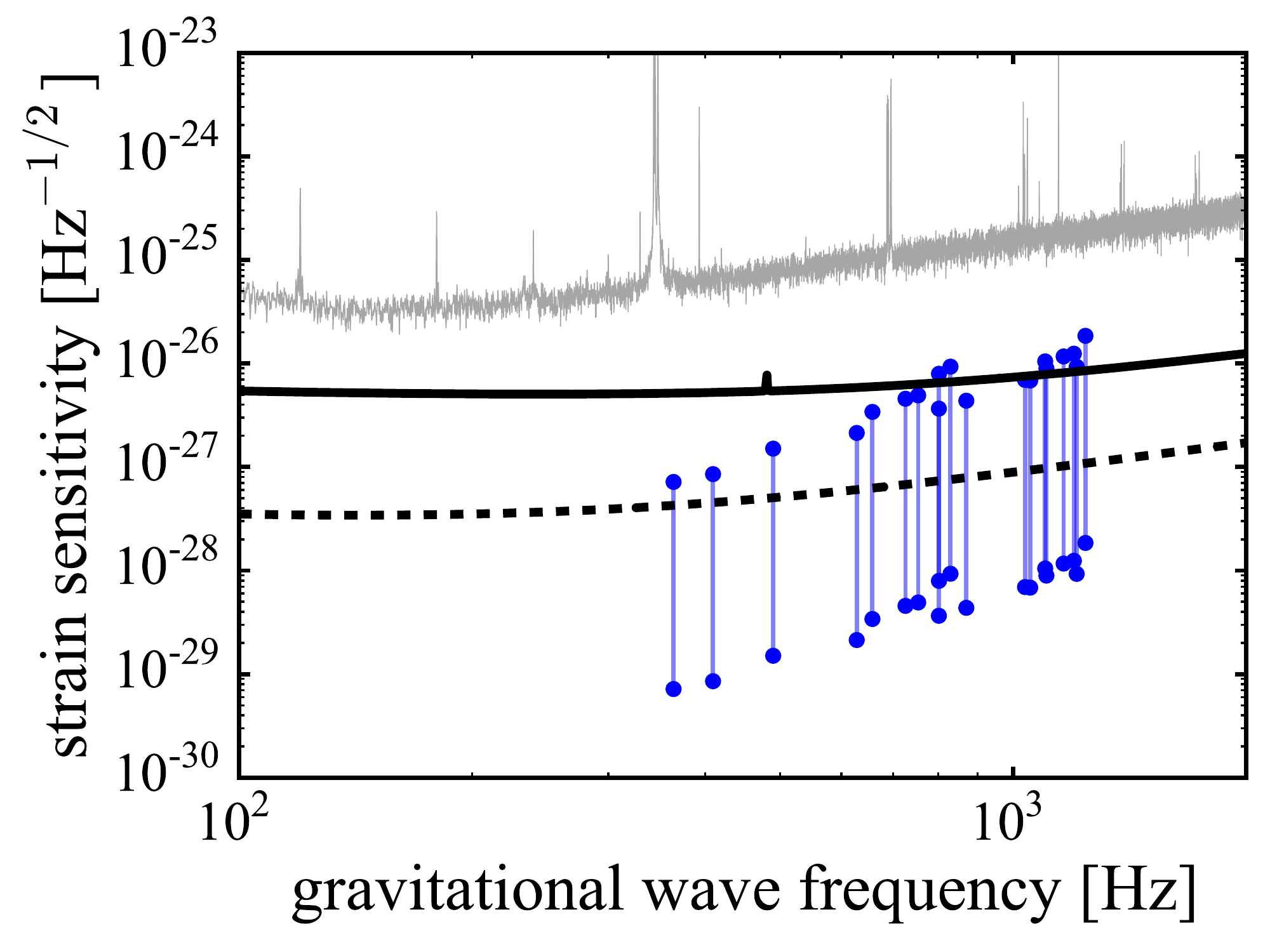}
	\caption{\label{Accrete} Left panel:  Gravitational wave torque balance limit for known accreting millisecond pulsars and systems with burst oscillations assuming gravitational wave emission at twice the neutron star spin period (blue dots).  Also shown is the torque balance limit for Sco X-1 (red curve) which has unknown spin period.  Data is collated from \citet{watts08} and \citet{haskell15}.  Right panel: Gravitational wave predictions for magnetic mountains on known accreting X-ray pulsars, where the range is for magnetic field strengths at the onset of accretion of between $B_\star=10^{10}$ and $10^{12}\,{\rm G}$ \citep[for details of the calculation, see][]{haskell15}.  In both panes, the solid and dashed black curves show the projected strain sensitivity for aLIGO and ET respectively, and the grey curve is the strain sensitivity for the initial S5 run, assuming two years of coherent integration time.  For comparison, current observational upper limits on Sco X1 from LIGO are $\lesssim8\times10^{-25}$ at $150\,{\rm Hz}$, which utilise a 10-day, semi-coherent analysis \citep{aasi15}.}
\end{figure*}

The torque balance limit for accreting systems with known spin periods is shown in the left hand panel of Figure \ref{Accrete} (blue dots), and for Sco X-1 (red curve) which has an unknown spin period.  The LIGO, aLIGO and ET sensitivity curves in this figure assume two years integration; see \citet{watts08} for more detailed, realistic estimates of signal-to-noise ratios of these systems.  Note that current upper limits of gravitational wave emission from Sco X-1 utilise a semi-coherent, 10-day search, which therefore yields less stringent upper limits than the LIGO curve in Figure \ref{Accrete} \citep{aasi15}.

\subsection{Unstable oscillation modes}\label{rmode}
Unstable oscillations modes, in particular $r$-modes, have received significant attention as potential sources of detectable gravitational waves.  These inertial modes -- toroidal modes of oscillation for which the restoring force is the Coriolis force -- are generically unstable to a phenomena known as the Chandrasekhar-Friedman-Schutz (CFS) instability \citep{chandrasekhar70,friedman78}.  Consider an oscillation mode in a rotating star.  The rotation induces both forward and backward propagating modes, but the backward propagating modes are being dragged forward by the star's rotation.  Although this retrograde mode continues to move backwards in the star's rotating frame, if the star is rotating sufficiently fast it will be prograde in the inertial frame.  The mode loses energy to gravitational waves, but carries with it positive angular momentum from the star.  This positive angular momentum is subtracted from the negative angular momentum of the mode, which becomes more negative, growing the amplitude of oscillation.  As the mode grows, it emits more positive angular momentum, and continues to grow even faster.  Hence, a sufficiently rapidly rotating star is unstable to gravitational wave emission.

At all stellar rotation rates, $r$-modes are retrograde in the comoving frame, but prograde in the inertial frame, implying they are generically unstable to the CFS instability \citep{andersson98,friedman98}.  A struggle therefore ensues between the gravitational waves which drive up the oscillation amplitude and the viscous damping that acts to suppress the mode amplitude.  There exists a narrow range in spin period and temperature for which gravitational wave emission dominates over viscous dissipation: at low temperatures, $T\lesssim({\rm few})\times10^9\,{\rm K}$, viscous dissipation is dominated by shear viscosity, while at high temperatures, $T\gtrsim({\rm few})\times10^9\,{\rm K}$, bulk viscosity is the main culprit.  Details of the $r$-mode instability window therefore depend sensitively on relatively unknown neutron star physics, including microphysics and complicated crust-core interactions \citep[for a review, see][]{andersson01}.  This somewhat simple picture describing the $r$-mode instability window is inconsistent with observed spins and temperatures of low-mass X-ray binaries, implying the complete physical picture behind this mechanism is ill-understood \citep{ho11b,haskell12}. 

\citet{strohmayer14} recently analysed RXTE observations of the accreting millisecond pulsar XTE J1751-305, finding evidence for a coherent oscillation mode during outburst that they attributed to either an $r$- or $g$-mode (for which buoyancy is the restoring force).  The $r$-mode interpretation is the most interesting in the context of gravitational wave emission; had the outburst occurred during the S5 LIGO run it would have been marginally observable \citep{andersson14}.  However, doubt has been cast over the $r$-mode interpretation; \citet{andersson14} also showed that the mode amplitude required to interpret the observations necessarily leads to large spin-down of the neutron star, which is inconsistent with the neutron star's observed spin evolution.

\subsection{Mountains}
On the other hand, permanent non-axisymmetries supported magnetically or elastically may generate a significant gravitational wave signature.  Accreted matter accumulates on the neutron star, is buried, compressed, and undergoes a range of nuclear reactions \citep[e.g.,][]{haensel90}.  In the non-magnetic case, asymmetries in the accretion lead to compositional and heating asymmetries, which induce stellar deformations.  Approximate expressions for the quadrupolar deformation can be derived in terms of the quadrupolar component of the temperature variation and reaction threshold energies \citep{ushomirsky00}.  Such deformations are limited by the maximum stress the crust can sustain before breaking \citep{ushomirsky00,haskell06,johnsonmcdaniel13}, which typically gives larger gravitational wave estimates than torque balance \citep{haskell15}.

Accretion also affects the structure of the neutron star's magnetic field which, in turn, changes the accretion dynamics.  The magnetosphere funnels accreted matter onto the poles, at which point it spreads towards the equator, dragging, compressing, and burying the local stellar magnetic field \citep{hameury83,melatos01}.  Such fields can lead to stellar deformations considerably larger than those due to the background magnetic field inferred from the external, dipole \citep{payne04,melatos05,vigelius09}, which may even generate gravitational waves detectable by aLIGO or the Einstein Telescope \citep{priymak11}, particularly if the buried field is $B\gtrsim10^{12}\,{\rm G}$ \citep{haskell15}.  This is shown in the right-hand panel of Figure \ref{Accrete}, where the gravitational wave signal from accreting systems with known spin periods is shown assuming initial magnetic fields of $B=10^{10}$ and $10^{12}\,{\rm G}$ \citep[for details of this calculation, see][]{haskell15}. Finally, it is worth remarking that, given a positive gravitational wave detection, an in-principle measurement of cyclotron resonant scattering features in the X-ray spectrum may be able to discern magnetic or elastic mountains \citep{priymax14,haskell15}.

\section{Pulsar Glitches}\label{glitch_sec}
Pulsar glitches are sudden jumps in the neutron star spin frequency with wide-ranging fractional amplitudes, $10^{-11}\lesssim\Delta\nu/\nu\lesssim10^{-4}$ \citep[][]{melatos08,espinoza11a,espinoza14}.  The exact mechanism driving a glitch is not fully understood, although it is clear that a sudden transfer of angular momentum occurs between the rapidly rotating superfluid interior and the outer crust \citep[for a recent review, see][]{haskell15b}.  Such large angular momentum transfer lends itself to the generation of gravitational radiation through a variety of avenues, as both broadband burst emission from the glitch and as a continuous wave signal during the glitch recovery phase.  

\subsection{Gravitational wave bursts}
Most theories of pulsar glitches rely on the general mechanism introduced by \citet{anderson75}.  The quantum mechanical nature of superfluids implies the neutron star's rotation is attributed to an array of $\sim10^{18}$ quantized superfluid vortices.  These thread the entire star, but are pinned to lattice sites and/or crustal defects, and hence are restricted from moving outwards as the crust spins down through the usual electromagnetic torques.  The superfluid core therefore retains a higher angular velocity than the crust of the star, and a differential lag builds up between these two components.  A glitch occurs when $\sim10^{7}$ -- $10^{15}$ vortices catastrophically unpin and move outwards, thereby rapidly transferring angular momentum to the crust.

That so many vortices are required to unpin simultaneously implies an avalanche trigger process must be at work.  Such an avalanche is likely non-axisymmetric, and hence capable of emitting gravitational waves through a variety of channels.  \citet{warszawski12} simulated the motion of vortices, showing that a burst of gravitational waves emitted through the current quadrupole has a characteristic strain
\begin{align}
	h_0\approx10^{-24}\left(\frac{\Delta\Omega/\Omega}{10^{-7}}\right)\left(\frac{\Omega}{10^2\,{\rm rad\,s}^{-1}}\right)^3\notag\\
	\times\left(\frac{\Delta r}{10^{-2}\,\rm{m}}\right)^{-1}\left(\frac{d}{1\,\mbox{kpc}}\right)^{-1},\label{glitch_h0}
\end{align} 
where $\Delta r$ is the average distance travelled by a vortex during a glitch.  The non-detection of gravitational waves from a glitch in the Vela pulsar during the fifth LIGO Science run in 2006 put an upper limit of $h_0\lesssim10^{-20}$ \citep{abadie11b}, implying an upper limit of $\Delta r\lesssim10^{-2}\,{\rm m}$ and a lower limit on the glitch duration $\gtrsim10^{-4}\,{\rm ms}$.

\begin{figure}
\includegraphics[width=1.0\columnwidth]{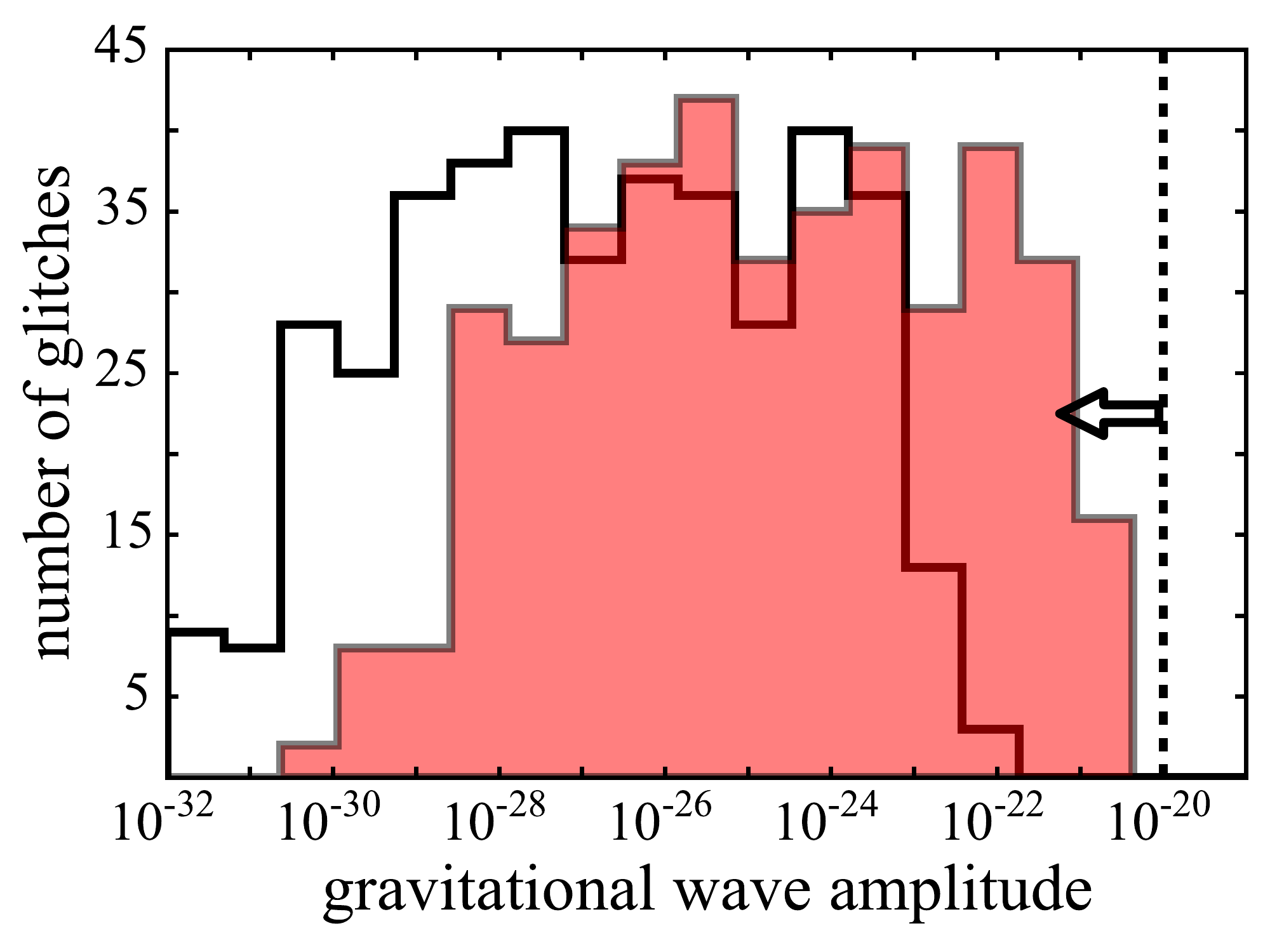}
	\caption{\label{glitch_fig}  Predicted gravitational wave amplitudes for known glitches using equation (\ref{glitch_h0}) with $\Delta r=10^{-2}\,m$ (black, unfilled histogram) and $\Delta r=10^{-4}$ (red, filled histogram).  The vertical, dashed line is the gravitational wave strain upper limit derived for the August 2006 glitch of the Vela pulsar during the LIGO S5 run \citep{abadie11b}}
\end{figure}

Figure \ref{glitch_fig} shows a histogram of estimates for the gravitational wave strain using equation (\ref{glitch_h0}), where the glitches are taken from the ATNF glitch catalogue \citep{manchester05}\footnote{\href{http://www.atnf.csiro.au/people/pulsar/psrcat/glitchTbl.html}{http://www.atnf.csiro.au/people/pulsar/psrcat/glitchTbl.html}}  The vertical dashed line shows the empirical gravitational wave strain upper limit from the August 2006 glitch of the Vela pulsar \citep{abadie11b}.

\subsection{Glitch recovery}
The motion of vortices is expected to excite hydrodynamic oscillation modes such as $f$- and $p$-modes, for which the pressure is the restoring force, and inertial $r$-modes \citep[e.g.,][]{andersson98a,andersson98,glampedakis09}.  Although $r$-modes are expected to be the dominant emission mechanism, recent estimates of their detectability with ground-based interferometers is pessimistic \citep{sidery10}.  More optimistic estimates have been provided for gravitational waves from meridional Ekman flows that are expected to couple the core and crust rotation post-glitch \citep{vaneysden08,bennett10}, although it is worth noting the model is sensitive to many unknown, dimensionless parameters such as the Ekman number, normalised compressibility and Brunt-V\"ais\"al\"a frequency.  Moreover, strong stratification may suppress Ekman flows to thin boundary layers in the outer regions of the star \citep{abney96,melatos12}, thereby reducing the gravitational wave strain estimates.  Finally, it has recently been suggested that vortex avalanches that are believed to trigger glitches leave behind long-lived, large-scale inhomogeneities on the vortex distribution that are potentially observable by aLIGO \citep{melatos15}.



\section{Magnetar Flares}\label{flare_sec}
Perhaps the most exotic of all neutron stars are the magnetars; isolated neutron stars with external dipole magnetic fields $B_{p}\gtrsim10^{14}$ G.  The decay of the immense magnetic field powers irregular bursts in hard X-rays and soft $\gamma$-rays, with typical peak luminosities $\sim10^{38}$---$10^{43}\,{\rm erg\,s}^{-1}$.  Three {\it giant} flares have been observed in our Galaxy\footnote{Interestingly, LIGO non-detections of gravitational waves from nearby short-duration, hard-spectrum gamma-ray bursts GRB 051103 \citep{abbott08d} and GRB 070201 \citep{abadie12b} rule out compact binary coalescences as their progenitors, implying these are most likely extragalactic magnetar giant flares.}, the largest of which was the extreme outburst from SGR 1806--20 on 27 December 2004 which emitted a total, isotropic energy of $2\times10^{46}$ erg \citep{palmer05}.  

Giant flares are commonly attributed to catastrophic rearrangements of the magnetic field, either internal or external to the neutron star  \citep[e.g.][]{duncan92,thompson95}.  Strong coupling between the magnetic field and the solid crust imply the latter will stress, and potentially rupture (although, see \citealt{levin12}), exciting oscillation modes in the star's crust and core.  Observations of quasi-periodic oscillations (QPOs) in the tails of two giant flares \citep{israel05, strohmayer05} has engendered excitement about the potential new field of neutron star asteroseismology.  In particular, the interpretation of QPOs as stellar magneto-elastic oscillations provide enticing potential to infer neutron star structural parameters \citep[see][and references therein]{levin06,glampedakis06,levin07, sotani07,sotani08b,colaiuda09,vanhoven11a,vanhoven11b,levin11,gabler13, gabler14,huppenkothen14a}. 

\subsection{KiloHertz gravitational waves}
First attempts to understand gravitational-wave emission from magnetar flares were optimistic.  \citet{ioka01} assumed an instantaneous change in the star's moment of inertia from a rearrangement of the global, internal magnetic field can excite the neutron star's fundamental $f$ mode (typically at frequencies, $f\sim1$--$2\,{\rm kHz}$), a study that was backed up by \citet{corsi11}.  These studies estimated that the energy in gravitational waves could be $E_\gw\sim10^{48}$--$10^{49}\,{\rm erg}$, comparable to the energy emitted in electromagnetic waves, and certainly detectable in the Advanced Detector Era.

In contrast, \citet{levin11} showed that, while the order-of-magnitude estimates of \citet{ioka01} and \citet{corsi11} are correct, in practice there is poor energy conversion into the $f$-mode, and prospects for gravitational wave detection in the Advanced Era are pessimistic.  The analytic calculation of \citet{levin11} looked at the energy conversion from a flare triggered in the star's magnetosphere.  \citet{zink12} and \citet{ciolfi12} performed complementary, general relativistic magnetohydrodynamic simulations using their respective codes \citep{lasky11,ciolfi11} for catastrophic reconfigurations of the internal magnetic field, also finding that the $f$-mode is not sufficiently excited to generate a detectable gravitational wave signature.  Although these works had different predictions for the gravitational wave scaling as a function of magnetic field strength, they both predicted approximately 10 orders of magnitude less energy being emitted in gravitational waves than \cite{corsi11}.

\begin{figure}
\includegraphics[width=1.0\columnwidth]{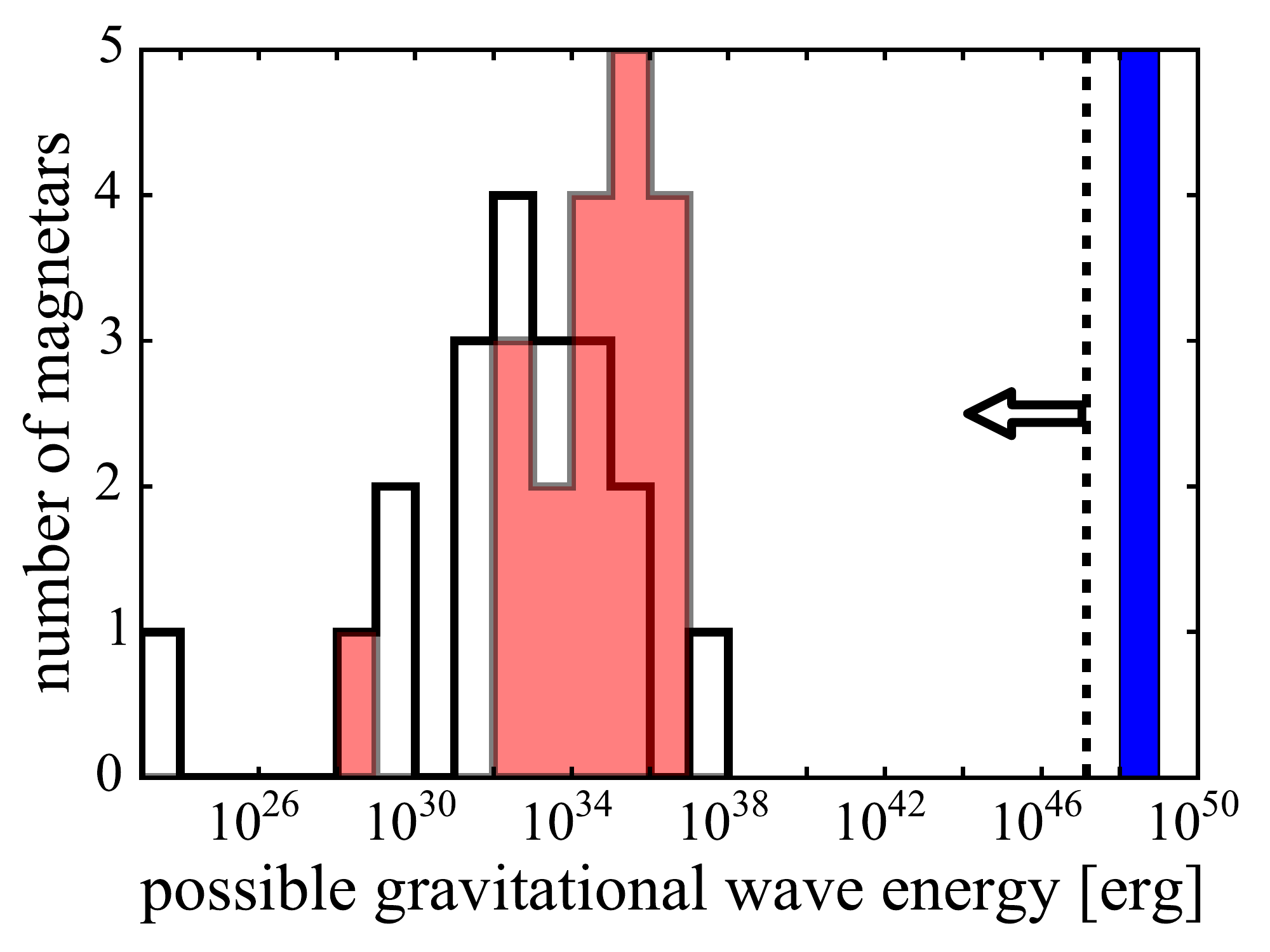}
	\caption{\label{giantflare_fig} Predictions of the possible gravitational wave energy emitted in $f$-mode oscillations if giant flares were to go off in each observed galactic magnetar given the calculations of \cite[][shaded red histogram]{ciolfi12} and \cite[][empty black histogram]{zink12,lasky12}.  Plotted in blue are the optimistic predictions of \cite{ioka01} and \citet{corsi11}, and the dashed black line is the upper limit on the $f$-mode gravitational wave energy from the giant flare from SGR 1806-20 \citep{abadie11}.}
\end{figure}

Figure \ref{giantflare_fig} shows a histogram of the possible gravitational wave energies emitted in the $f$-mode from known galactic magnetars\footnote{The known magnetars are taken from the McGill Magnetar catalogue \citep{olausen14}.} using the calculations of \cite{ciolfi12} (shaded red histogram) and \cite{zink12,lasky12} (empty black histogram).  In blue are the gravitational wave predictions from \cite{ioka01} and \cite{corsi11}, while the dashed black line gives the observed upper limit on the $f$-mode gravitational wave energy from the 2004 giant flare in SGR 1806-20 \citep[][see below]{abadie11}.

\subsection{Low-frequency gravitational waves}
Although the $f$-mode couples well to gravitational radiation, one of the reasons it is not an optimal source for ground-based detectors is that it is damped through the emission of gravitational radiation in $\lesssim0.1\,{\rm s}$ \citep{detweiler75,mcdermott88}.  Longer lasting, lower frequency modes excited from magnetar flares have been suggested as potential sources of detectable gravitational waves \citep{kashiyama11,zink12,lasky12}.  To generate sufficient mass motions, and therefore a detectable gravitational wave signal, these modes need to be global core oscillations, and hence are unlikely the cause of the observed QPOs.  A key uncertainty for these low frequency modes is their damping time, which underpins their gravitational wave emission.  If these modes form an Alfv\'en continuum of frequencies, where the QPOs arise as the edges or turning points of the continuum \citep{levin07}, then resonant absorption should quickly redistribute energy, resulting in short lifetimes ($\lesssim1\,{\rm s}$) for the QPOs \citep{levin07,levin11,gabler11,vanhoven12,huppenkothen14a}, and pessimistic estimates for gravitational wave emission.  On the other hand, if global Alfv\'en waves are excited in the core \citep{zink12} then they may live for 100's of seconds, and be potentially detectable by third-generation interferometers, or even aLIGO in the case of extremely strong ($\sim10^{16}\,{\rm G}$) magnetic fields in the core \citep{glampedakis14}.

\subsection{Current observational limits}
At the time of the giant flare from SGR 1806-20, the $4\,{\rm km}$ Hanford detectors was the only interferometer operating.  An upper limit of $E_\gw<7.7\times10^{46}\,{\rm erg}$ was achieved around the $92.5\,{\rm Hz}$ QPO, with a significantly weaker constraint placed in the kHz range where one expects the $f$-mode.  The most sensitive gravitational wave search of the giant flare was from the AURIGA bar detector \citep{baggio05}.  They searched for exponentially decaying signals (with decay time $0.1\,{\rm s}$) in a small frequency band around $900\,{\rm Hz}$.  Subsequent LIGO/Virgo searches of regular flares from magnetars have yielded comparable gravitational wave limits to the original 2004 burst \citep{abbott08,abadie11}.

While galactic giant flares only occur approximately once per decade, the regularity of magnetar flare storms implies they may provide an attractive alternative.  If such flares also excite normal stellar modes, it is easy to imagine that stacking gravitational wave data from multiple flares could allow for sufficient integration time.  Indeed, such a gravitational wave search has been designed \citep{kalmus09} and implemented to search for gravitational waves in LIGO/Virgo data \citep{abbott09b}.  These attempts have further been buoyed by the successful detection of QPOs in ordinary magnetar flares \citep{huppenkothen14b,huppenkothen14c}.

\section{Superfluid Turbulence}\label{turb_sec}
Soon after the realisation that neutron star matter should form a superfluid condensate \citep[][and references therein]{baym69}, \citet{greenstein70} suggested this superfluid should be in a highly turbulent state\footnote{The beautifully succinct abstract is seldom seen today: `The neutron superfluid in most neutron stars should be in a highly turbulent state.  If so, this turbulence drastically alters its rotational properties' \citep{greenstein70}.}.  Such turbulence can be either hydrodynamic 
\citep[e.g., ][]{peralta05, peralta06a, peralta06b, peralta08} or quantum mechanical 
\citep[e.g.,][]{tsubota09,andersson03a,mastrano05,andersson07,link12a,link12b}.  Regardless, the turbulence is likely driven by differential rotation between the superfluid core (with Reynolds number ${\rm Re}\sim10^{11}$) and the solid crust that is being slowly spun down by electromagnetic torques.

\subsection{Individual sources}
Although the turbulence is axisymmetric when averaged over long times, it is instantaneously non-axisymmetric, and therefore emits stochastic gravitational waves.  The signal was first calculated in \citet{peralta06a} by numerically solving the spherical Couette-flow problem with ${\rm Re}\lesssim10^{6}$.  A subsequent analytic calculation \citep{melatos10} made a more robust prediction of the gravitational wave amplitude and spectrum, showing in particular that the peak of the gravitational wave signal occurs near the inverse of the turbulence decoherence timescale, $\tau_c$.  For reasonable neutron star parameters, this decoherence timescale is 
\begin{align}
	\tau_c\approx26\left(\frac{\Delta\Omega}{10\,{\rm rad\,s}^{-1}}\right)^{-1}\,{\rm ms},
\end{align}
where $\Delta\Omega$ is the difference in angular frequency between the crust and the core of the neutron star, which can potentially be related to the star's angular frequency, $\Omega$, through the Rossby number, ${\rm Ro}=\Delta\Omega/\Omega$.
For many pulsars, this places the peak of the gravitational wave signal in LIGO's most sensitive band, although it is worth noting there is considerable uncertainty in $\Delta\Omega$.  Finally, the root-mean-square of the gravitational wave strain is
\begin{align}
	h_{\rm rms}\approx8\times10^{-28}\left(\frac{\Delta\Omega}{10\,{\rm rad\,s}^{-1}}\right)^3\left(\frac{d}{1\,{\rm kpc}}\right)^{-1}.
\end{align}

\begin{figure}
\includegraphics[width=1.0\columnwidth]{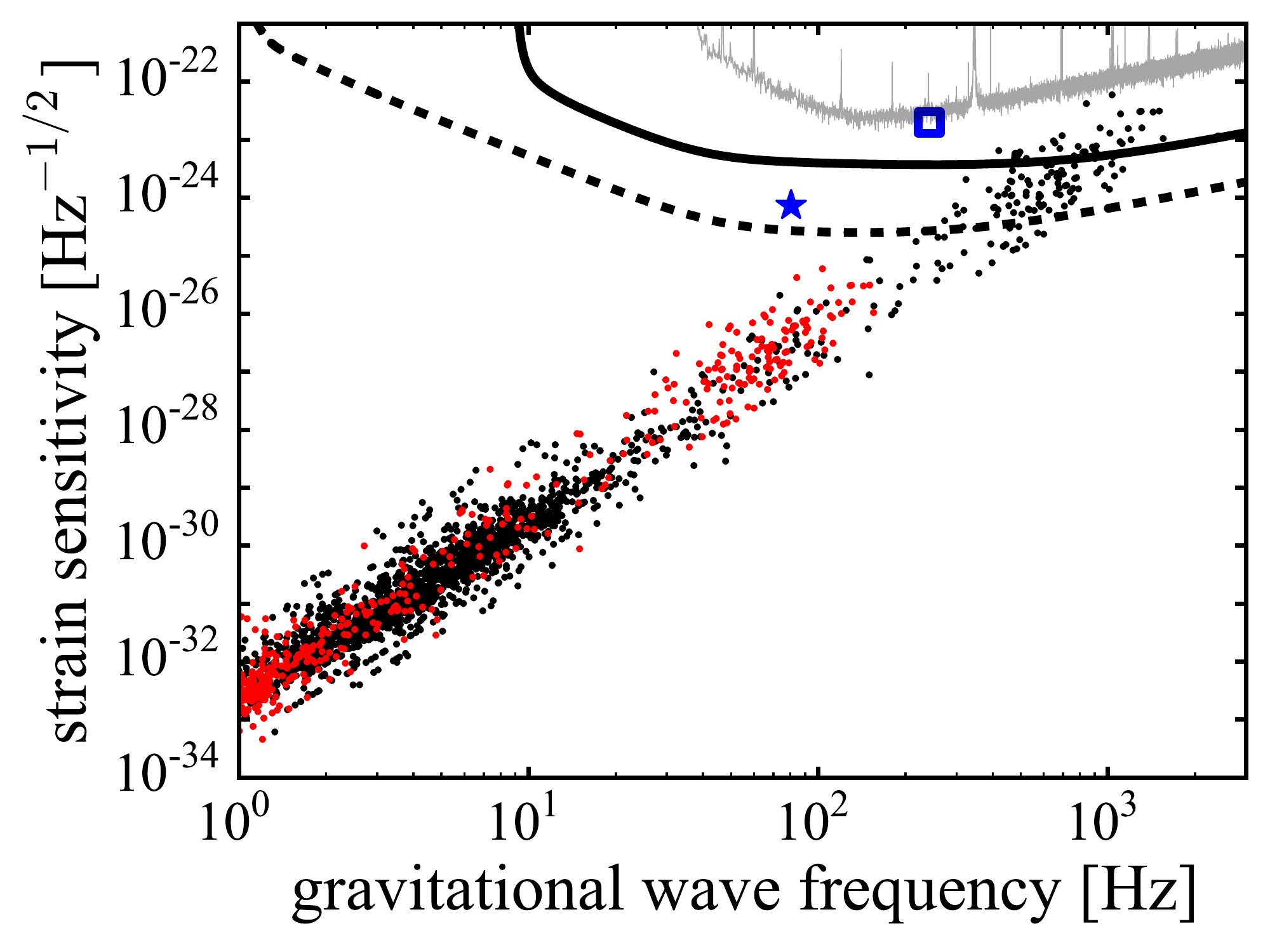}
	\caption{\label{turbulence_fig} Peak gravitational wave amplitudes from superfluid turbulence from galactic pulsars with ${\rm Ro}=\Delta\Omega/\Omega=10^{-1}$ (black points) and $10^{-2}$ (red points).  The blue star and square are hypothetical, nearby ($d=10\,{\rm pc}$) rapid rotators with spin periods $p=3$ ms and 10 ms, and ${\rm Ro}=10^{-2}$ and $10^{-1}$, respectively.}
\end{figure}

Figure \ref{turbulence_fig} shows the peak gravitational wave predictions for the superfluid turbulence model for observed galactic pulsars and millisecond pulsars with ${\rm Ro}=10^{-1}$ (black points) and $10^{-2}$ (red points).  It is worth stressing that these values of the crust-core shear, $\Delta\Omega$, are most likely exaggerated; in the same paper \cite{melatos10} showed that ${\rm Ro}\lesssim10^{-2}$ for typical millisecond pulsars.  Calculations of vortex \citep{seveso12} and flux-tube \citep{link03b} pinning suggest $\Delta\Omega$ should vary, but will generally be $\lesssim10^{-2}$.  One therefore concludes that turbulence excited in newly born protoneutron stars, or extremely close rapid rotators, are the only possible source for Advanced Era gravitational wave interferometers.  The blue square and star in Figure \ref{turbulence_fig} show the peak gravitational wave emission for hypothetical neutron stars situated at $d=10\,{\rm pc}$ with spin periods $p=3$ ms and $10$ ms, and Ro$=10^{-2}$ and $10^{-1}$, respectively, the former of which would be detectable by aLIGO.

\subsection{Stochastic background}
A cosmological population of neutron stars with turbulent cores produces a stochastic gravitational wave background that peaks in aLIGOs most sensitive frequency band \citep{lasky13a}.  Although there are large uncertainties in the expected amplitude of the background due to a lack of understanding of $\Delta\Omega$, the shape of the spectrum is relatively unique, with it being well-approximated by a piecewise power-law $\Omega_\gw(f_\gw)=\Omega_{\alpha}f_\gw^{\alpha}$, with $\alpha=7$ and $\alpha=-1$ for $f_\gw<f_c$ and $f_\gw>f_c$, respectively.  Here, $f_c$ is the population-weighted average of $1/\tau_{\rm c}$, and $\Omega_\gw(f_\gw)$ is the energy density in the gravitational wave background as a fraction of the closure energy density of the Universe. 

It is worth noting that turbulent convection in main sequence stars also produces a gravitational wave signal \citep{bennett14}.  Interestingly, the loudest gravitational wave signal detectable on Earth may come from the Sun as, for frequencies $f_\gw\lesssim3\times10^{-4}\,{\rm Hz}$, the Earth lies in the Sun's near zone, in which the wave strain scales significantly more steeply with distance, $h_{\rm rms}\propto d^{-5}$ \citep[cf. $\propto d^{-1}$ in the far zone;][]{cutler96,polnarev09}.  This gravitational wave signal is most relevant in the micro- to nHz regime, and therefore most applicable to pulsar timing experiments.  However, the gravitational-wave scaling at low-frequencies is uncertain as Kolmogorov scaling for turbulence breaks down below $f_\gw\lesssim10^{-8}\,{\rm Hz}$.

%

\section{Conclusions}
Ground-based gravitational wave interferometers are already contributing to our understanding of interesting astrophysical phenomena.  But non-detections will only progress the field so far; the first direct detection of gravitational radiation will herald a new scientific field of study, and will allow new insight into the most exotic regions of our Universe.  

While the first detection with aLIGO is expected to be from the inspiral and merger of a compact binary system, there are many unknown, and ill-understood, mechanisms that can generate gravitational waves with significant amplitudes from isolated neutron stars.  This review highlights many of those mechanisms where we have some understanding of the key, physical processes.  It also highlights the vast uncertainty of many of these predictions, showing that they rely on knowledge of neutron star physics beyond current capabilities.  But this is what makes the field so fascinating; a positive detection of gravitational waves from any of the mechanisms discussed in this article will allow an unprecedented view into the heart of neutron stars, where some of the most exotic physics in the Universe takes place.

\begin{acknowledgements}
PDL is supported by the Australian Research Council Discovery Project DP140102578.  I am extremely grateful to Bryn Haskell, Daniel Price, Kostas Glampedakis, and Karl Wette for useful comments on the manuscript.
\end{acknowledgements}

\bibliographystyle{apj}
\bibliography{Bib}

\end{document}